\documentclass[%
 reprint,
 amsmath,amssymb,
 aps,
 twocolumn,
]{revtex4-2}

\usepackage{printlen}
\usepackage{color,soul}
 
\usepackage{xcolor}
\usepackage{graphicx}
\usepackage{dcolumn}
\usepackage{bm}
\usepackage{hyperref}
\usepackage[mathlines]{lineno}%
\usepackage{subfigure}
\usepackage[utf8]{inputenc}
\usepackage[T1]{fontenc}

\newcommand{\matr}[1]{\mathbf{#1}}
\newcommand*{\affaddr}[1]{#1} 
\newcommand*{\affmark}[1][*]{\textsuperscript{#1}}
\begin{document}

\preprint{APS/123-QED}

\date{\today}

\title{Affective and interactional polarization align across countries}

\author{Max Falkenberg\affmark[1,*], Fabiana Zollo\affmark[2,3], Walter Quattrociocchi\affmark[4], J{\"u}rgen Pfeffer\affmark[5],  and Andrea Baronchelli\affmark[1,6,*]
\begin{center}
\affaddr{\affmark[1]{ \textit{City University of London, Department of Mathematics, London EC1V 0HB, UK}}} \\
\affmark[2]{ \textit{Department of Environmental Sciences, Informatics and Statistics, Ca’ Foscari University of Venice, Venice, Italy}} \\
\affmark[3]{ \textit{The New Institute Centre for Environmental Humanities, Venice, Italy}} \\
\affmark[4]{ \textit{Department of Computer Science, Sapienza University of Rome, Rome, IT}} \\
\affmark[5]{ \textit{School of Social Science and Technology, Technical University of Munich, Germany}} \\
\affaddr{\affmark[6]{ \textit{The Alan Turing Institute, British Library, London NW1 2DB, UK}}} \\
\affaddr{\affmark[*]{Corresponding authors: max.falkenberg@city.ac.uk, abaronchelli@turing.ac.uk}}
\end{center}
}
\begin{abstract}
\vspace{0.5cm}
\textbf{\noindent 
Political polarization plays a pivotal and potentially harmful role in a democracy. However, existing studies are often limited to a single country and one form of polarization, hindering a comprehensive understanding of the phenomena. Here we investigate how affective and interactional polarization are related across nine countries (Canada, France, Germany, Italy, Poland, Spain, Turkey, UK, USA). First, we show that political interaction networks are polarized on Twitter. Second, we reveal that out-group interactions, defined by the network, are more toxic than in-group interactions, meaning that affective and interactional polarization are aligned. Third, we show that out-group interactions receive lower engagement than in-group interactions. Finally, we show that the political right reference lower reliability media than the political left, and that interactions between politically engaged accounts are limited and rarely reciprocated. These results hold across countries and represent a first step towards a more unified understanding of polarization.
}
\end{abstract}
\maketitle

\section*{Introduction}
Political polarization plays a critical and yet not fully understood role in the effective running of a democracy. Some polarization is important for driving debate in public policy, improving the deliberation of ideas and, arguably, facilitating access to diverse information sources across groups \cite{iandoli2021impact}. However, severe polarization can stifle debate and drive animosity between groups which, in the worst cases, can lead to violence \cite{mccoy2018polarization}. 

Given these factors, a vast literature has emerged aiming to better understand political polarization. Scholars place a particular emphasis on affective polarization, defined as the tendency to dislike ones partisan opponents, given that it may undermine the mechanisms which allow a democracy to function \cite{iyengar2019origins}. 
Researchers have shown that affective polarization has grown steadily in the USA \cite{boxell2022cross}. While the exact reasons for this increase are unclear, research has linked affective polarization with social identity theory \cite{iyengar2019origins} and partisan sorting \cite{roccas2002social,ojer2023charting}. Political parties are increasingly associated with specific social identities and demographics \cite{mason2015disrespectfully,mason2018uncivil},  increasing the perceived distance between partisans and driving out-group animosity \cite{levendusky2016mis}. 

Research on affective polarization has traditionally been carried out using survey data. However, the rise of the internet has seen social media emerge as an alternate public for the study of polarization \cite{bail2022breaking,lee2014social}. On social media, the most commonly studied form of polarization is interactional polarization \cite{yarchi2021political} - sometimes referred to as structural \cite{salloum2022separating} or social network polarization \cite{tokita2021polarized} -  which looks at how the interaction patterns between ideological groups are segregated.  

The role of social media in driving polarization is disputed \cite{nordbrandt2021affective,guess2023reshares,guess2023social, kubin2021role, lorenz2023systematic}; many argue that polarization on social media simply mirrors the underlying polarization of our societies. However, the richness and availability of social media data has made it an invaluable forum for studying the mechanisms of polarization \cite{bail2018exposure,flores2022politicians, cinelli2021echo,tornberg2022digital}, depolarization \cite{chen2021polarization,xia2022russian}, its evolution over time \cite{waller2021quantifying,falkenberg2022growing}, and for testing potential interventions and countermeasures \cite{guess2023reshares}.  Additionally, social media allows for a direct measurement of partisan animosity through the analysis of inter-group messages (e.g., using toxicity analysis, or dictionaries of polarized language \cite{simchon2022troll}). Twitter, now rebranded as X, is particularly important for polarization research given its outsized influence on politicians \cite{stieglitz2013social} and journalists \cite{hu2012breaking,molyneux2022legitimating}. Several studies have shown that political Twitter networks are structurally polarized \cite{bovet2019influence,flamino2023political}. However, the majority of studies on interactional polarization do not consider how these structures relate to out-group animosity. 

While the impact of social media on affective polarization remains a matter of debate, it is clear that these platforms have lowered the barriers to out-group interaction, thereby enabling the evolution of negative out-group emotion into toxic out-group messaging. Worryingly, this may in fact increase engagement with social media \cite{rathje2021out}. For this reason, it is critical to understand how polarized social network structures relate to affective polarization. There is some evidence that interactional polarization aligns with affective polarization, for instance in relation to US elections \cite{grimminger2021hate, saveski2021structure}, the US far-right \cite{mekacher2023systemic}, UK politicians \cite{agarwal2021hate}, in relation to Covid-19 \cite{cinelli2021dynamics,miyazaki2022aggressive}, and following violent events in Israel \cite{yarchi2021political}. However, these examples remain limited to individual countries and contexts. 

Recent work highlights why we should be careful not to assume that findings on polarization generalize between regions: Across 12 OECD countries, the USA has seen the largest increase in affective polarization over the last four decades, with 6 countries showing a decrease in affective polarization \cite{boxell2022cross}. Similarly, recent studies have found conflicting outcomes regarding the effect of deactivating social media on political polarization in the USA and in Bosnia-Herzegovina \cite{allcott2020welfare,asimovic2021testing}. Despite this, the majority of studies, particularly in the most prestigious journals \cite{flamino2023political,rathje2021out,bail2018exposure,waller2021quantifying,robertson2023users,guess2023reshares,guess2023social,nyhan2023like,gonzalez2023asymmetric,voelkel2023interventions,finkel2020political}, still only consider polarization in the United States.

One of the reasons for the lack of cross-country studies on social media is the difficulty acquiring sufficiently large datasets which are not keyword specific \cite{roozenbeek2022democratize}. Here we overcome this limitation by using a uniquely complete Twitter dataset which includes all public interactions across a 24 hour period \cite{pfeffer2023just}. Coupled with a second dataset of known elected politicians on Twitter \cite{van2020twitter}, we are able to study how affective polarization aligns with interactional polarization across nine countries (Canada, France, Germany, Italy, Poland, Spain, Turkey, UK, USA) covering seven languages.   

In the remainder of this paper, we first give an overview of the datasets studied and visualize the  network of politicians on Twitter. Then, we compute the spectrum of interactional polarization, showing that it broadly aligns with a left-right political dimension (with the exception of Germany, where the primary divide is establishment-populist). Grouping users on each side of the structural divide, we show that for all nine countries out-group interactions are more toxic, but receive lower engagement, than in-group interactions. We also show that the political right consistently reference lower reliability media outlets than the political left. Finally, we reveal that only a minority of interactions from politically engaged accounts are with other politically engaged accounts, that these interactions share a common ally-enemy structure across political groups, and that interactions between partisans are rarely reciprocated. We end by contextualizing our work and discussing its implications for the wider study of political polarization.

\section*{Results}
To study polarization across countries on Twitter, we use a complete dataset of all Twitter posts (including retweets) from a 24 hour period in September 2022 (see Methods), totaling 375 million tweets. In this paper we are interested in political interactions. To identify these, we use a second dataset of known elected politicians from 26 different countries \cite{van2020twitter} to label all content involving politically engaged Twitter users. This includes all posts authored by politicians, all interactions with those politicians (excluding likes), and all posts and interactions by the accounts who at any point have engaged with these politicians. From this filtering process, we identify nine countries where there is sufficient engagement with politicians across the 24 hour period (at least 5000 unique user pairs between politicians and their retweeters \footnote{This threshold is determined experimentally. For countries with an insufficient number of unique user pairs between politicians and retweeters, the computed latent ideology is highly dependent on the Twitter interactions of a small number of highly active accounts.  The resulting distribution of ideology scores is often not representative of the diverse range of political views found in a country.}) to enable a robust comparison of affective and interactional polarization. These countries are Canada, France, Germany, Italy, Poland, Spain, Turkey, the United Kingdom, and the United States. The filtered dataset of political Twitter interactions is broken down in detail in the SI. In its totality, the filtered dataset includes the interactions of 140 thousand unique users with 1,837 unique elected politicians across the 24 hour period.

\subsection*{Visualizing politicians on Twitter}
To start, we visualize the network of political Twitter interactions to gain an intuitive understanding of the structure of multi-national political communication.
\begin{figure*}[t]
    \centering
\includegraphics{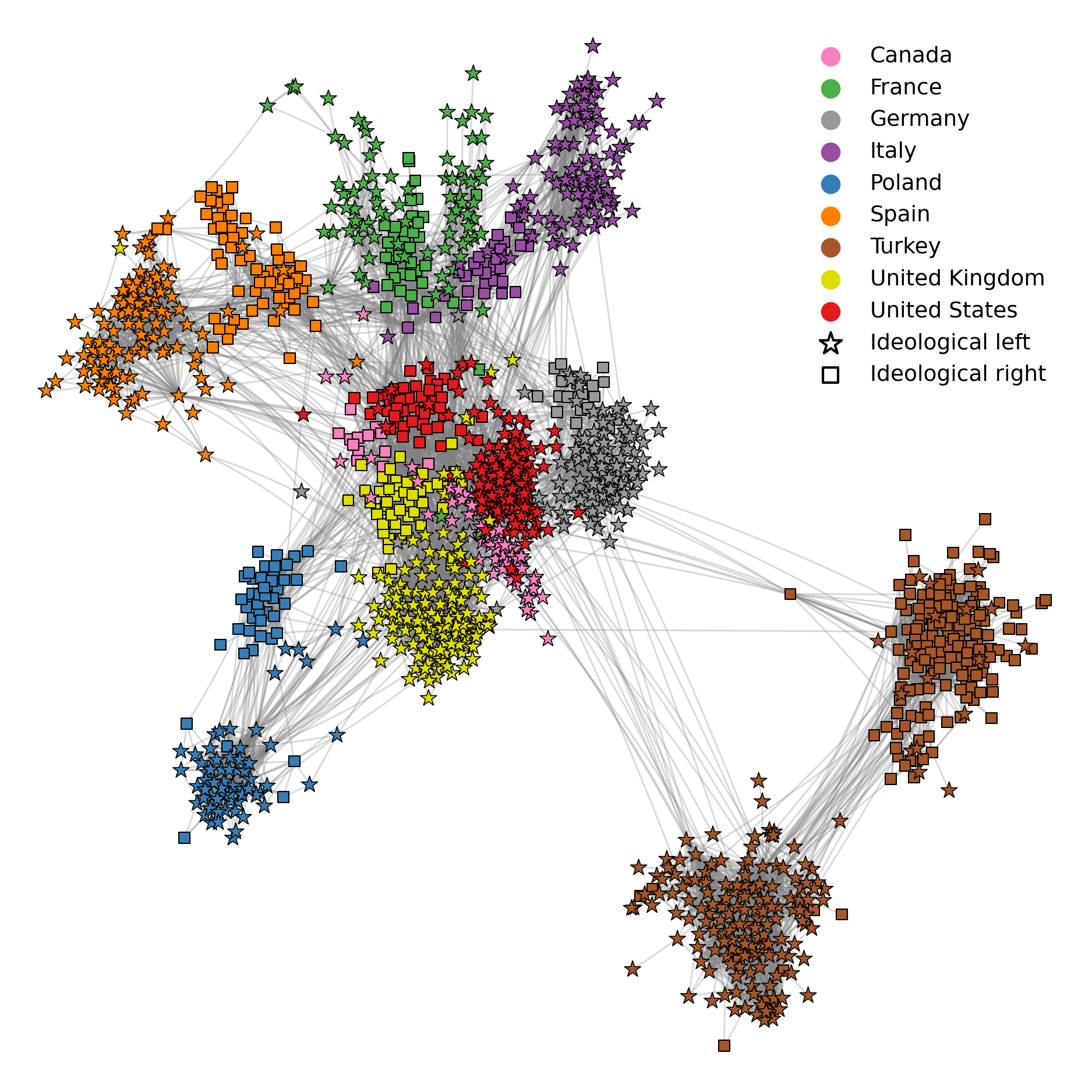}
    \caption{\textbf{The network of political Twitter interactions is segregated across countries and polarized within then.} The figure depicts a co-retweet network where nodes correspond to individual politicians and an edge is drawn between two nodes if those politicians share a common retweeter (see Methods). Square nodes correspond to the ideological right, star nodes correspond to the ideological left, as determined using the latent ideology (see Fig.~\ref{fig:fig2}). The network visualization is produced using a force directed drawing algorithm (see Methods).}
    \label{fig:fig1}
\end{figure*}

Since our aim is to identify politicians who are ideologically aligned our polarization analysis uses retweet networks, a common approach in many Twitter-based polarization studies \cite{falkenberg2022growing,flamino2023political,salloum2022separating}. We focus on retweets since they are generally evidence of a Twitter user endorsing the message of the original poster \cite{metaxas2015retweets}, as opposed to other Twitter interactions (mentions, quotes or replies) which may indicate a positive, negative or neutral relationship between the two users. Retweets also uniquely refer to a single user as opposed to replies and mentions where multiple users may be referenced. Using the retweet framework, if two politicians share a large number of common retweeters they likely share similar ideological views (see for example \cite{falkenberg2022growing}). 

Figure~\ref{fig:fig1} shows the co-retweet network of elected politicians from the nine countries studied. Each node in the network corresponds to a single elected politician, colored according to country. Two politicians are connected by an edge if they share at least two common retweeters. To avoid spurious connections, we exclude a small number of retweets from highly active accounts (possibly spam) who engage with many politicians. 

From the network visualization we make three qualitative observations: (1) The political Twitter discussion in each country is largely self-contained, separate to the political discussion from other countries. (2) Within each country, there is a clear separation between the political left and right, forming well defined clusters. (3) Across the nine countries studied, Anglophone countries (US, UK, Canada) are at the center of the political retweet network. These countries also exhibit some overlap of their political factions: The US left (right) is structurally closer to the Canadian left (right), than to the US right (left); in the SI we provide evidence that these countries share a larger fraction of common users than country pairs which do not share the same language. We quantify the lack of transnational political communication in the SI where we show that very few users interact with politicians from more than one country.  

\subsection*{Universal interactional polarization across countries}
To formalize our observation of polarized political Twitter networks, we now measure the spectrum of interactional polarization in each of the nine countries studied. For each country, we construct a bipartite network between the country's elected politicians active on Twitter (the ``influencers''), and all remaining Twitter users who retweet those politicians (the ``retweeters''). Connections between politicians, and between retweeters, are not required to compute the spectrum of interactional polarization and are ignored in the current analysis.

From each bipartite network, we compute a one-dimensional spectrum of ideological scores using the latent ideology method, originally developed for follower networks in \cite{barbera2015birds}, adapted to retweet networks in \cite{flamino2023political}, and applied using elected politicians as the set of influencers in \cite{falkenberg2022growing}. A precise mathematical formulation for the latent ideology is provided in the Methods. Intuitively, the method produces a one-dimensional ordering where Twitter users who retweet similar sets of politicians are close to each other in the ordering. Exact ideological scores produced are arbitrary and should not be compared across networks. Here, we rescale the derived ideological scores so that the two dominant peaks of the ideology distribution align with scores of $-1$ and $+1$ respectively. 

\begin{figure*}[p!]
    \centering
\includegraphics{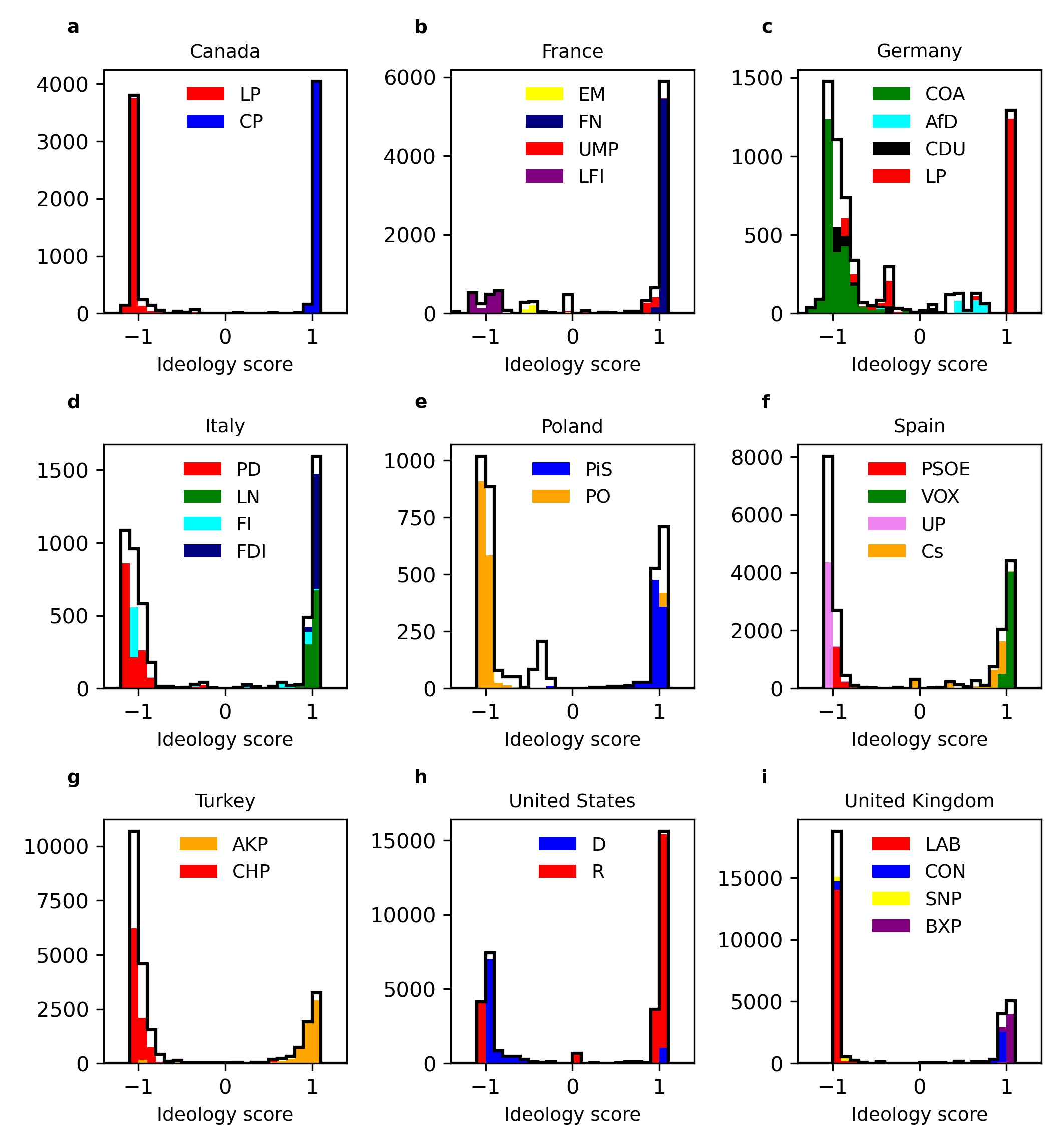}
    \caption{\textbf{Political interaction networks are structurally polarized in each of the nine countries studied.} For each country, we compute the latent ideology of Twitter users based on their retweet interactions with elected politicians (see Methods). In each subfigure, the histogram outlined in bold shows the number of Twitter users with an ideology score in the binned range. Colored bars show the modal political party of users in the binned ideology range. Area in white corresponds to users without a unique modal political party or interacting with other political parties. \scriptsize  
    Canada: \{LP: Liberal Party, CP: Conservative Party\}. 
    France: \{EM: En Marche, FN: Rassemblement National, UMP: Les R\'epublicains\}.
    Germany: \{COA: Coalition (Sozialdemokratische Partei Deutschlands / Freie Demokratische Partei / B\"undnis 90/Die Gr\"unen), AfD: Alternative f\"ur Deutschland, CDU: Christlich Demokratische Union Deutschlands / Christlich-Soziale Union in Bayern, LP: Die Linke\}. 
    Italy: \{PD: Partito Democratico, LN: Lega Nord, FI: Forza Italia, FDI: Fratelli d'Italia\}.  
    Poland: \{PiS: Prawo i Sprawiedliwo\'s\'c, PO: Platforma Obywatelska\}. 
    Spain: \{PSOE: Partido Socialista Obrero Espa\~nol, VOX: Vox, UP: Podemos, Cs: Ciudadanos\}. 
    Turkey: \{AKP: Adalet ve Kalkınma Partisi, CHP: Cumhuriyet Halk Partisi\}.
    United States: \{D: Democrats, R: Republicans\}. 
    United Kingdom: \{LAB: Labour, CON: Conservative, SNP: Scottish National Party, BXP: Brexit Party / UK Independence Party / Reform\}.
    }
    \label{fig:fig2}
\end{figure*}

The distribution of user ideology scores for each country is shown in Fig.~\ref{fig:fig2}. The histogram is shaded according to the modal political party of the politicians retweeted by users in the binned range of ideology scores. Users who do not retweet a unique political party, or whose modal retweeted party received little engagement, are not shaded.

Figure~\ref{fig:fig2} shows that political Twitter interactions are polarized in each of the nine countries, with a bimodal (or multi-modal) distribution of ideology scores. In general, retweeters who align with a specific political party are found in only one of the two dominant peaks in the ideology distribution of each country, not both. For example, in Canada (Fig.~\ref{fig:fig2}a) retweeters who align with the left-leaning Liberal party are found in the left peak with ideology scores $<0$. Conversely, retweeters who align with the right-leaning Canadian Conservative party are found in the right peak with ideology scores $>0$. Similarly, in the USA, most users who align with the Democrats are found in the left peak, whereas most users who align with the Republicans are found in the right peak. In Poland, we find the center-right Platforma Obywatelska party (PO; Civic Platform) in the left peak and the populist-right Prawo i Sprawiedliwość party (PiS; Law \& Justice) in the right peak. Both parties are on the political right, but in a relative sense PO is further left than PiS. 

The only case where the latent ideology does not align with the left-right dimension is Germany, where the primary structural divide is along the establishment-populist dimension. Users who retweet politicians from political parties in the governing coalition have ideological scores less than zero, whereas most users from the Left party (LP) and the far-right Alternative for Germany (AfD) have ideological scores larger than zero. This merger is discussed in the SI and relates to the unified AfD/LP position criticizing the German government's position on the Russia-Ukraine war.

Based on these observations, in the following we refer to the ideological left as all influencers (politicians) and retweeters with an ideology score less than 0, and the ideological right as all politicians and retweeters with an ideology score larger than 0. Note, however, that references to the left-right political spectrum are used loosely; individual countries have their own political nuances. 

There are some cases where the retweeters of a political party do not align with the rest of their party in the interaction network. The best example of this is in the US (Fig.~\ref{fig:fig2} panel h) where a number of users labeled as Republicans are shown on the left, and a number of users labeled as Democrats are shown on the right. This reflects users who retweet party outliers. In the case of the US, most users whose modal party is Republican but have an ideological score less than zero are retweeters of Liz Cheney (an elected Republican), whereas most users whose modal party is Democrat but have an ideological score greater than zero are retweeters of Tulsi Gabbard (an elected Democrat; defected to the Republicans in October 2022, after our data collection period). Both politicians are known outliers from the dominant position of their parties, and their structural alignment with opposition parties on Twitter has been noted previously in work studying the US far-right \cite{mekacher2023systemic}. An overview of political outliers in other countries is provided in the SI.

\subsection*{Out-group interactions are more toxic than in-group interactions}
We have shown that in each of the nine countries studied the network of political Twitter interactions is structurally polarized, in most cases along a broadly left-right spectrum. We now ask whether this spectrum of interactional polarization aligns with affective polarization, referring to out-group hostility \cite{finkel2020political}.

As we have seen, defining groups according to political party is limiting due to the presence of party-outliers (e.g., Liz Cheney and Tulsi Gabbard in the US). Hence, we use an interactional approach, defining an ``in-group'' interaction as any Twitter mention where both users (the mentioner and mentionee) are from the same ideological group, i.e., both from the left (scores $<0$) or right (scores $>0$). Conversely, we refer to a Twitter interaction as ``out-group'' when the two users are from opposed groups (one left, one right). We acknowledge that this grouping may oversimplify the political reality of some countries, but consider it an acceptable approximation for the current study; for alternative approaches to studying polarization in multi-party contexts see \cite{martin2023multipolar,baumann2021emergence,peralta2023multidimensional}. 
\begin{figure}[h!]
    \centering
    \includegraphics{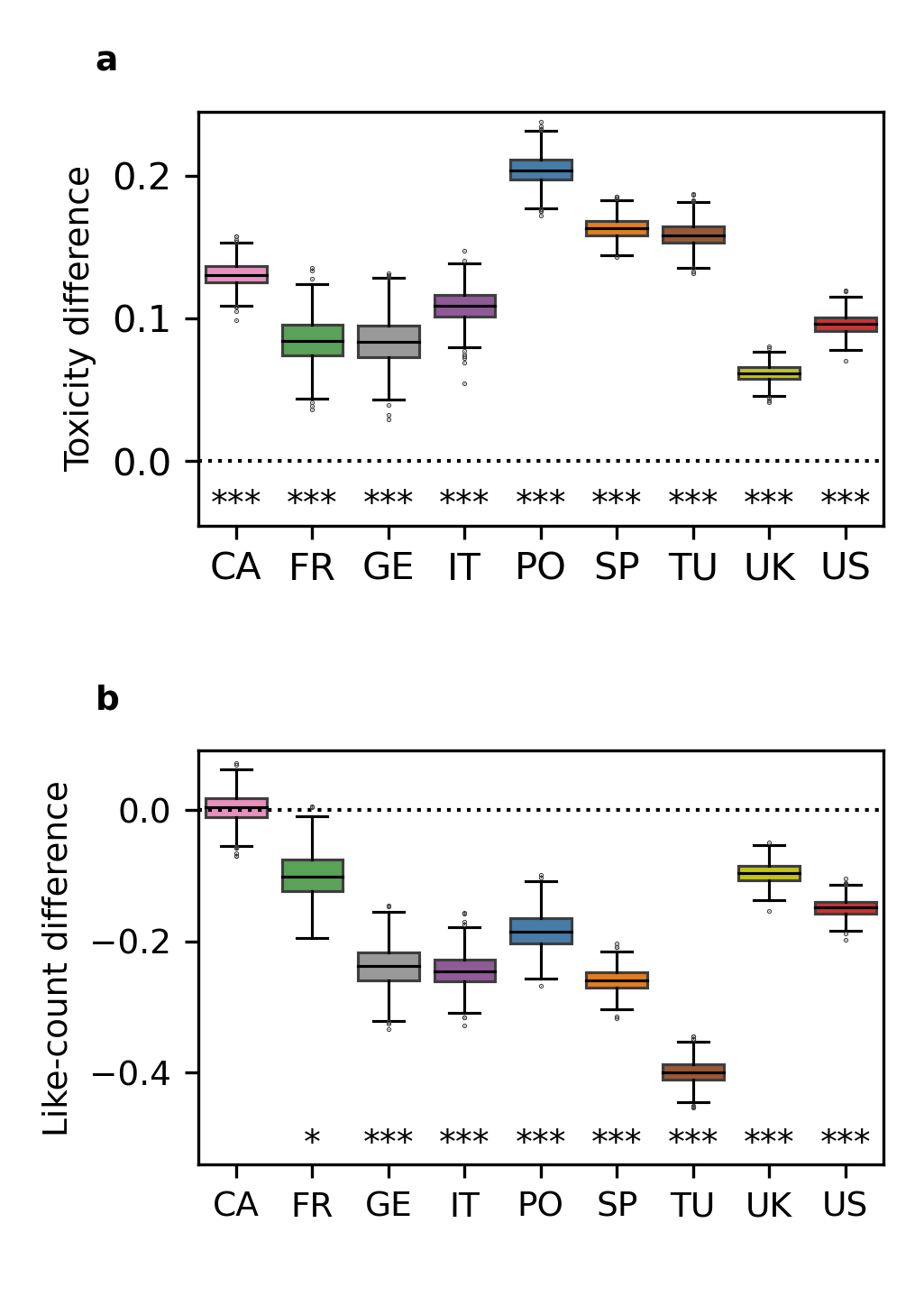}
    \caption{\textbf{Out-group interactions are more toxic, but receive lower engagement, than in-group interactions.} (a) Boxplots for the bootstrapped difference between the median out-group interaction toxicity, less the median in-group interaction toxicity for each of the nine countries. (b) Boxplots for the bootstrapped difference between the mean log likes received by an out-group post relative to the mean log likes received by an in-group post. Boxplot show the bootstrapped median, interquartile range (IQR), whiskers for $1.5$ times the IQR from the hinge, and points for outliers (see Methods). Stars indicate statistical significance using the non-parametric Mann-Whitney U test on full distribution (see Methods): $p<0.05$: *, $p< 0.01$: **, $p<0.001$: ***.}
    \label{fig:fig3}
\end{figure}

To measure affective polarization, we calculate the toxicity of original posts (not retweets) which include a mention between ideologically labeled users. For English, French, Italian and Spanish language posts we compute toxicity scores twice, once using Google Perspective API \cite{perspective}, and a second time with Detoxify using BERT sentence classifiers as a robustness check (see SI). For German and Polish posts we compute toxicity scores with Google Perspective API only; these languages are not compatible with Detoxify. For Turkish we compute toxicity scores with Detoxify only; Turkish is not compatible with the Perspective API. In the SI, we show that our results are robust using either toxicity model. 

The derived toxicity scores from both models fall in the range $[0,1]$. Posts with scores near $0$ are the least likely to be considered toxic by a human labeler. Conversely, posts with scores near $1$ are the most likely to be considered toxic by a human labeler. To ease comparison between countries, in the following we analyze posts according to their toxicity quantile rather than raw toxicity score.

Figure~\ref{fig:fig3}(a) shows boxplots of the bootstrapped difference in the median toxicity quantile of out-group interactions less the median toxicity quantile of in-group interactions for each country. Comparing the toxicity distributions using the non-parametric Mann-Whitney U test (see Methods), out-group interactions are significantly more toxic than in-group interactions. This result shows that in each of the nine countries studied, out-group interactions, defined based on the interaction network, are more toxic than in-group interactions, confirming that affective and interactional polarization align.

Our results are robust if we only consider posts authored by the political left, or by the political right (see SI). For all nine countries, the political left are more toxic when interacting with the political right than when interacting with the political left (statistically significant in each case). Similarly, the political right are more toxic when interacting with the political left than when interacting with the political right in eight of the nine countries (statistically significant in each case). The UK is the only outlier where the political right are slightly more toxic when interacting with the right than when interacting with the left; this is likely due to toxic interactions between the UK right (Conservative party) and far-right (e.g., Brexit party, Reform party).

Comparing the out-group interactions of the political left to the out-group interactions of the political right, we do not find a consistent trend suggesting that the political left are more toxic than the political right, or vice versa, across countries  (see SI).

\subsection*{Out-group interactions receive lower engagement than in-group interactions}
Having found that affective polarization aligns with interactional polarization, we now ask whether out-group interactions receive lower engagement than in-group interactions. Figure~\ref{fig:fig3}(b) shows boxplots of the bootstrapped difference in the mean log like-count ($\text{log}_{10}[\text{likes} + 1]$; likes recorded 10 minutes after a post first appeared online, see Methods) received on posts with an out-group mention and the mean log like-count received on posts with an in-group mention. The log like-count is used to avoid a small number of posts with very large engagement dominating the mean; the $+1$ avoids errors due to posts with zero likes.
The panel shows that out-group posts receive lower engagement than in-group posts (result not statistically significant in Canada). This result is robust if we only consider posts authored by the political left or the political right (see SI). Across all nine countries, posts authored by the political left mentioning another user from the political left receive higher engagement than posts mentioning a user from the political right (statistically significant in each case). For the political right,  posts mentioning another user from the political right receive higher engagement than posts mentioning a user from the political left in seven of the nine countries (no significant difference in Canada and France).

These results suggest that out-group interactions receive lower engagement than in-group interactions. Consequently, given the realities of an attention economy, this is arguably an incentive for Twitter users to prioritize in-group interactions over out-group interactions.

\begin{figure}[h!]
    \centering
\includegraphics{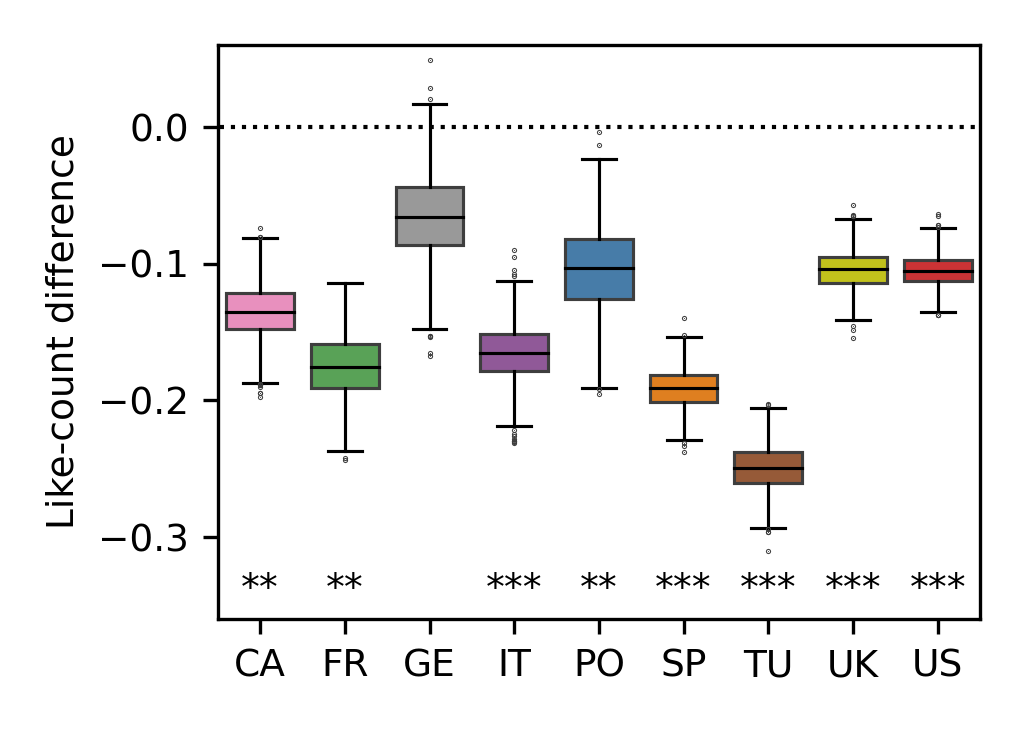}
    \caption{\textbf{High toxicity interactions receive lower engagement than low toxicity interactions.} Boxplots show the bootstrapped mean log like-count received by high toxicity posts, minus the mean log like-count received by low toxicity posts. High toxicity interactions are defined as the top 10\% most toxic posts in each country. 
    Boxplot show the bootstrapped median, interquartile range (IQR), whiskers for $1.5$ times the IQR from the hinge, and points for outliers (see Methods).
    Stars indicate statistical significance using the non-parametric Mann-Whitney U test on full distribution (see Methods): $p<0.05$: *, $p< 0.01$: **, $p<0.001$: ***.}
    \label{fig:fig4}
\end{figure}

One possible reason for lower out-group engagement may be the confounding factor of post toxicity. Figure~\ref{fig:fig4} shows boxplots for the bootstrapped mean engagement (log-likes) received by the most toxic posts in each country, less the mean engagement received by lower toxicity posts. The figure shows that for all nine countries posts with high toxicity receive lower engagement than low toxicity posts (result for Germany not statistically significant). This may be an authentic reflection of Twitter-users behavior, showing that users are less likely to interact with toxic posts. However, it is also possible that this difference is the result of content moderation policies reducing the visibility of, or removing, offensive posts. 

\subsection*{The political right reference lower reliability media sources than the political left}
We now assess the news domains shared by the political left and right on Twitter. To do so, we use news media reliability classifications provided by NewsGuard (see Methods). This provider is chosen since their classifications cover news outlets in the USA, UK, Canada, Germany, France and Italy; other providers typically only provide classifications for English language news outlets. We do not have access to news media reliability classifications for Spain, Poland and Turkey. 

Figure~\ref{fig:fig4} shows that, for the six countries where NewsGuard provides media reliability scores, the political right reference to lower reliability news media sources than the political left. This confirms and extends the results from a previous study of US elites which showed that conservative ideology correlates with higher exposure to unreliable news \cite{mosleh2022measuring}, but extends this to a wider set of countries.
\begin{figure}[h!]
    \centering
    \includegraphics{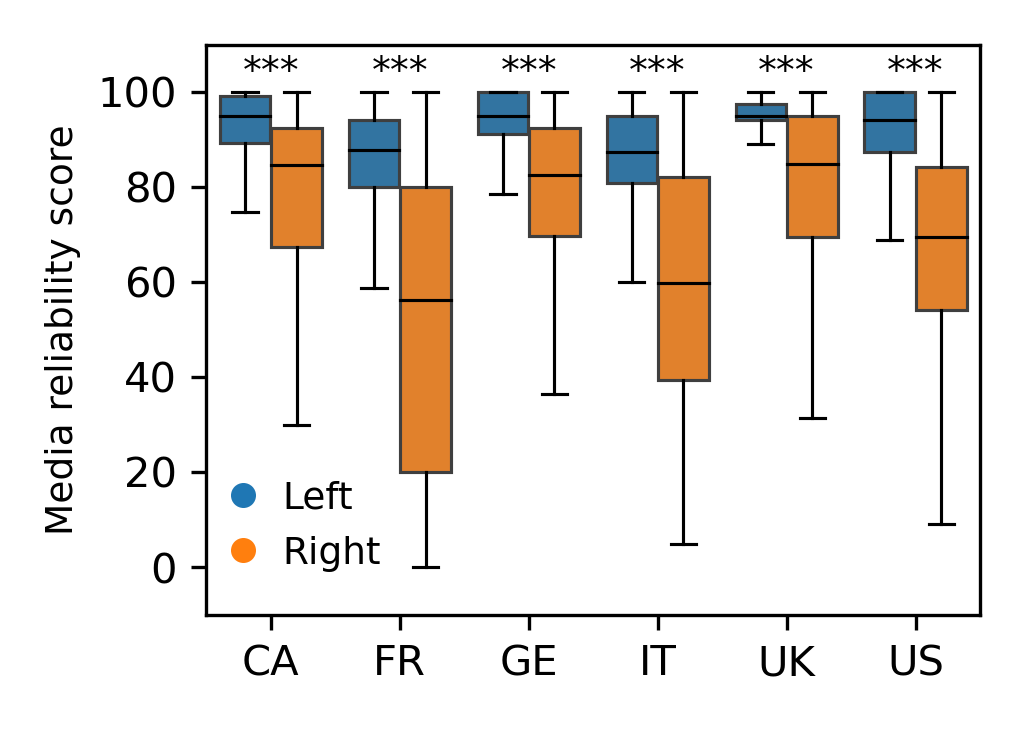}
    \caption{\textbf{Media reliability scores are lower for the ideological right than the ideological left.} Boxplots of the media reliability scores corresponding to news domain URLs found in Twitter posts authored by the ideological left and right respectively. Reliability scores are provided by NewsGuard (see Methods); data does not cover Poland, Spain or Turkey. Boxplot show the bootstrapped median, interquartile range (IQR), whiskers for $1.5$ times the IQR from the hinge, and points for outliers (see Methods). Stars indicate statistical significance using the non-parametric Mann-Whitney U test (see Methods): $p<0.05$: *, $p< 0.01$: **, $p<0.001$: ***.}
    \label{fig:fig5}
\end{figure}

\subsection*{The majority of Twitter interactions are with apolitical accounts}
\begin{figure*}[t]
    \centering
    \includegraphics{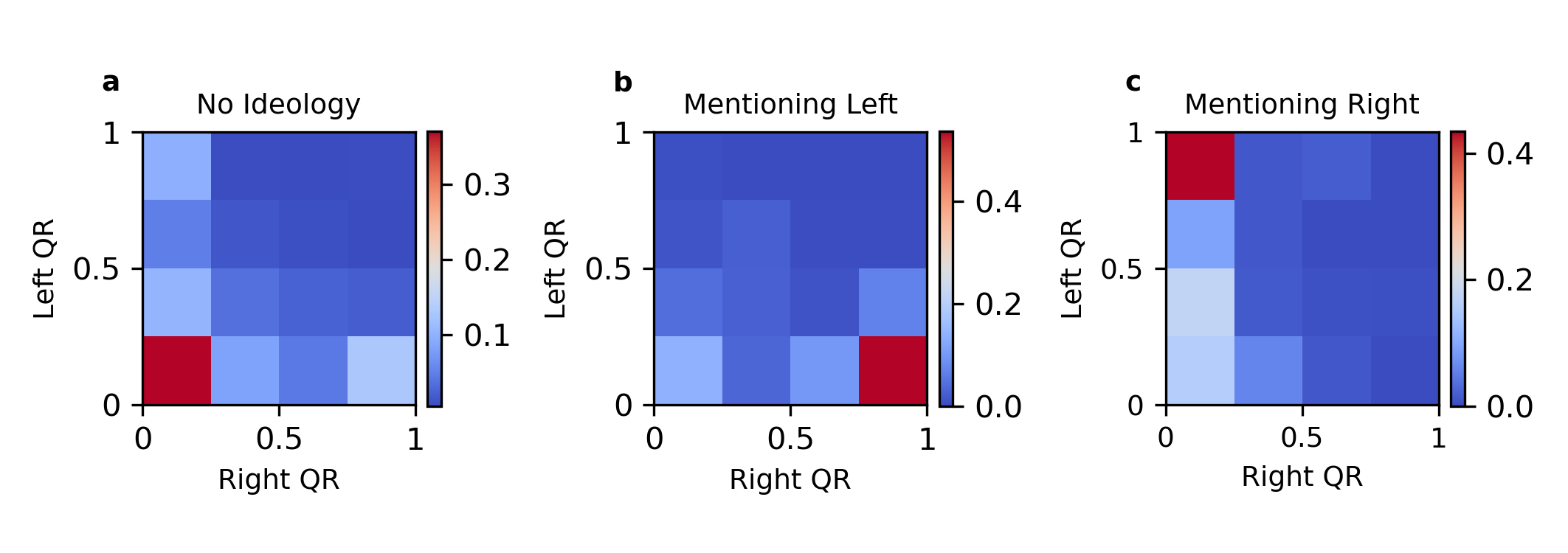}
    \caption{\textbf{The political left and right differentiate between allies and enemies in their interaction patterns, but treat apolitical accounts equally.} Binned quote-ratios (QR) for users mentioned who (a) do not have a political ideology, (b) have a left-leaning ideology, and (c) have a right-leaning ideology. The quote-ratio is computed twice for each mentioned group; once with users from the ideological left as the mentioners (y-axis: Left QR) and once with the right as the mentioners (x-axis: Right QR). Subpanels show the mean computed across all nine countries equally weighted; individual countries in SI. }
    \label{fig:fig6}
\end{figure*}

Our focus thus far has been on political Twitter interactions. However, there is substantial evidence to suggest that social media is primarily used for non-political purposes; most social media users are not politically engaged \cite{wojcieszak2022most}. Here, we study the differences in how politically engaged Twitter users interact with each other, as opposed to with non-politically engaged users. 

To achieve this, we compute the ``quote-ratio'', a metric of cohort-level endorsement defined as the ratio between the number of times a fixed cohort of ideologically aligned users mention a given account in an original tweet (quote tweet, original tweet, or reply), normalized by the number of times the account is mentioned by the same cohort in any tweet (including retweets).  This metric was developed in \cite{mekacher2023systemic} based on the premise that retweets are generally indicative of an endorsement on Twitter \cite{metaxas2015retweets}, whereas non-retweet mentions can be used in a positive, negative, or neutral manner. Hence, if a cohort of ideologically aligned users frequently mention, but never retweet, an account, then the members of the cohort likely disagree with the views of the mentioned account. In contrast, given that retweets are more common than other interaction types, accounts which are disproportionately retweeted are generally seen as endorsed by members of the cohort. The efficacy of this metric is demonstrated in \cite{mekacher2023systemic} where the US far-right are shown to have a low quote-ratio when mentioning Republican politicians and far-right media sources, and a high quote-ratio when mentioning Democrat politicians and left leaning media sources.

Figure~\ref{fig:fig6} shows the quote-ratio computed using the ideological left as the mentioning cohort (vertical axis) and the ideological right as the mentioning cohort (horizontal axis), broken down according to the group of users mentioned. Figure~\ref{fig:fig6}(a) shows the quote-ratio for mentioned users who do not engage with elected politicians on Twitter and are therefore not assigned an ideological score. Figures~\ref{fig:fig6}(b) and~\ref{fig:fig6}(c)  show the quote-ratio for mentioned accounts from the ideological left and right respectively. Each panel is averaged across the nine countries studied, with each country equally weighted (individual countries are shown in the SI).

The figure shows that, when interacting with accounts who are not politically engaged (panel a), most accounts are disproportionately retweeted (i.e., endorsed) by both the ideological left and right. These apolitical interactions represent the majority of interactions in each country (between 89-96\%; see SI). In contrast, accounts from the left (panel b) have a low quote-ratio when mentioned by other accounts from the left (i.e., the left endorse the left), but a large quote-ratio when mentioned by accounts from the right (i.e., the right mention, but do not endorse, the left). The reverse is also true, with mentioned accounts from the right (panel c) having a low quote-ratio when mentioned by other accounts from the right, but a high quote-ratio when mentioned by the left. 

This pattern is largely robust at the individual country level, see SI, revealing the common ally-enemy structure of political interactions across countries. Mentions of the political right in France and Germany are slight outliers. In the case of Germany, this is likely because the main structural divide observed in Fig.~\ref{fig:fig2}(c) is between establishment and populist parties, not right and left. In the case of France, this is likely due to relatively low activity from the French left leaning parties (see Fig.~\ref{fig:fig2}(b)). 

\subsection*{Out-group interactions are rarely dyadic}
Our analysis has not considered whether interactions are unidirectional or dyadic (i.e., if user A mentions user B, does user B then mention user A?). In the SI, we show that across all countries dyadic interactions are rare. On average, for in-group user pairs, 8.2\% of interactions are dyadic, whereas for out-group interactions 2.2\% are dyadic. For every country, in-group interactions are more likely to be dyadic than out-group interactions. This demonstrates that, across countries, out-group conversation with political opponents, as opposed to unidirectional broadcasting, is rare. 

A natural question is whether there is a difference in the toxicity of unidirectional interactions as opposed to dyadic interactions. Unfortunately, we do not observe enough dyadic interactions to allow for a robust analysis of dyadic toxicity in the current study. This question should be investigated in future work. 

\section*{Discussion \& Conclusion}
We have shown that affective and interactional polarization align across countries and languages on Twitter: When dividing a country's political interaction network into its two primary structural groups, out-group interactions are more toxic than in-group interactions. These results are robust for all nine countries and for both the political left and the political right.
This is an important step forward in building a global understanding of how different types of political polarization are related.  

We have addressed the multi-faceted nature of polarization by drawing on the strengths of social media research, while also employing the insights from political science. Specifically, our analysis defines the ideology of Twitter users through their partisan association, identified via their endorsement of elected politicians. Our results show how the supporters of a given political party typically cluster in a single group, structurally separated from their political opponents. However, our results also reveal how partisan non-conformists are essentially treated as members of the political opposition. Despite sharing few political views, Liz Cheney was heavily retweeted by users who align with the Democratic party. Conversely, Republicans interacted with Liz Cheney as if she were a Democrat (Republican commentators have referred to her as a ``RINO'': Republican in name only). This behavior is observed across a number of countries and raises the worrying prospect that, online, there is no political middle ground.  

Going beyond the focus on out-group animosity, our social media lens shows how out-group interactions generally draw lower engagement than in-group interactions. The extent to which this is an authentic user-driven result, or an artifact of content down-ranking by the Twitter recommendation system (for structural reasons, because the content is toxic, or otherwise) remains unclear. However, it does suggest an incentive for politically engaged users to prioritize engagement from a single political group as opposed to from a politically diverse user base. This reinforcing cycle may in turn worsen interactional polarization over time.

As well as including partisan messages and mentions, political Twitter posts often reference a range of different media sources, as is expected given the polarized media landscape \cite{wojcieszak2016partisan}. Our results confirm previous work demonstrating the difference in the reliability of media sources shared by partisans on the political left and right \cite{mosleh2022measuring}, but extends this result to include the USA, UK, France, Italy, Germany and Canada. In each case, we find that the political right reference lower media reliability outlets than the political left.

Our primary focus has been on political Twitter interactions. However, politically engaged Twitter users interact with apolitical users as well as with other partisans. Focusing on this divide, we have shown how partisan's differ in their interactions with other partisans, as opposed to with individuals who are not politically engaged. We find that there is a common ally-enemy structure in how the political left and right interact with each other, and that interactions with apolitical accounts are structurally similar. These represent the majority of the interactions in our dataset, confirming and extending previous work for the USA which showed that most Twitter users are not politically engaged \cite{wojcieszak2022most}. 

Finally, we show that despite the prevalence of, often toxic, out-group communication, these interactions are rarely reciprocated. This supports the echo chambers theory of social media \cite{cinelli2021echo} in that engaged partisans do not appear willing to engage in active conversation with political opponents. Whether this is a positive or a negative is unclear; there are arguments to suggest that cross-party communication may in fact reduce affective polarization \cite{levendusky2021we}. However, it also comes at the risk of increased exposure to toxic content and hate speech, and the negative consequences that an individual experiences as a result. 

There are limitations to our study which present opportunities for future work. First, the countries we study are largely Western and developed. While the importance of these countries should not be underestimated, there remain open questions as to  whether our results generalize to other regions, particularly to countries in the global south. Second, the study focuses exclusively on Twitter (now X). Future work should consider a similar analysis on other platforms. However, we stress that understanding polarization on Twitter remains critically important: It is one of the most influential social media sites for politicians and journalists \cite{molyneux2022legitimating,stieglitz2013social}, and results for Twitter will likely have some relevance for Twitter's emerging competitors (e.g., Threads, Bluesky) which use similar interaction mechanisms. Third, our study uses data gathered during a single 24 hour observation window. While this ensures coherent results from the slow changes in political discourse, it also raises the possibility that results may differ across larger time windows. There is evidence that the structure of political interactions evolve to reflect the changing political landscape of a country (for example in Pakistan \cite{baqir2023social}). However, the feasibility of such long-term studies is now in question given the recent restrictions to academic access for social media data \cite{roozenbeek2022democratize}. Finally, our study is observational in nature. Our results do not point towards a causal relationship between observed interactional polarization and observed affective polarization. However, our research emphasizes that studying polarization in a siloed manner may be counterproductive since the different forms of polarization may have interdependent mechanisms. 

In summary, our findings contribute towards a unified understanding of how different forms of polarization are related, and how these results generalize across countries. 
Within this picture, our work provides context for future work looking at how to reduce partisan animosity which, if left untackled, may lead to political violence \cite{esau2023destructive}.

\section*{Methods}
\subsection*{Data}
The Twitter data analyzed was acquired by Pfeffer \emph{et al.} \cite{pfeffer2023just}. The dataset includes all public Twitter posts across a 24 hour period starting on September 21, 2022. Posts were downloaded almost exactly ten minutes after they appeared online. Data is publicly available in the form of tweet IDs at \cite{pfeffer2023data}; the authors of \cite{pfeffer2023just} acknowledge that following changes to the Twitter API, downloading these tweets using the tweet IDs provided may be difficult. Consequently, the authors encourage anyone interested in the dataset to contact them for collaboration.

\subsection*{Network visualization}
The network visualization in Fig.~\ref{fig:fig1} is a co-occurence network of elected politicians in the nine countries studied. Each node corresponds to an elected politician who was active during the 24 hour period. For visual clarity, we only show politicians who were retweeted by two unique users in the 24 hour observation window. Two politicians (nodes) are connected by an edge if they are both retweeted by the same user. When constructing the network, we remove edges which are due to highly active accounts, defined as accounts who have retweeted over ten different politicians. This limits the number of spurious edges in the network which may be due to automated accounts spamming retweets. Finally, for visualization purposes we remove any politicians who are not part of the giant connected component. The resulting network is drawn manually with a layout derived using ForceAtlas2 \cite{jacomy2014forceatlas2}, with nodes colored according to country. The shape of nodes is determined by their ideological score as computed using the latent ideology, see Fig.~\ref{fig:fig2}.

\subsection*{Latent ideology}
Ideological scores for each country's political Twitter network are derived using the latent ideology method developed in \cite{barbera2015birds}. We use the same adaptation applied in \cite{falkenberg2022growing} for use with a bipartite representation of Twitter retweet networks between a set of $m$ politicians (influencers) and their $n$ retweeters (users). Influencer-influencer connections and user-user connections are ignored when constructing the bipartite network. 

We start with an $n \times (m+1)$ matrix $\matr{A}$ where each element $a_{ij}$ is the number of times user $i$ has retweeted politician $j$. For each country, we include all known elected politicians who were active in the 24 hour period and were retweeted at least once. In column $m+1$ we include a ``dummy politician'' who is retweeted by every user (i.e., a columns of $1$s) to ensure that the bipartite network is a single connected component. This dummy politician is removed from the analysis once ideological scores have been derived. 

Once the matrix $\matr{A}$ has been constructed, we compute the matrix normalized according to the number of retweets  as $\matr{P}=\matr{A}(\sum_{ij}a_{ij})^{-1}$, where the vector of row and column sums are given respectively by $\matr{r}=\matr{P}\matr{1}$ and $\matr{c}=\matr{1}^T\matr{P}$, and using the matrices $\matr{D}_r=\text{diag}(\matr{r})$ and $\matr{D}_c=\text{diag}(\matr{c})$. From this, we can compute the matrix of standardized residuals of the adjacency matrix as $\matr{S}=\matr{D}^{-1/2}_r(\matr{P}-\matr{r}\matr{c})\matr{D}^{-1/2}_c$.
This residual matrix accounts for differences in activity of retweeters and differences in the popularity of individual politicians (i.e., how often each politician is retweeted).
Next, single value decomposition is applied to the matrix $\matr{S}$ as
$\matr{S}=\matr{U}\matr{D}_{\alpha}\matr{V}^T$ with $\matr{U}\matr{U}^T=\matr{V}\matr{V}^T=\matr{I}$ and $\matr{D}_\alpha$ being the singular values diagonal matrix.
The ideological scores of the retweeted users is given by the standard row coordinates $\matr{X}=\matr{D}_r^{-1/2}\matr{U}$. In line with previous studies, we only consider the first dimension that corresponds to the largest singular value. 
We rescale the ideological scores of retweeters such that the two largest peaks in the ideology score distribution align with scores of $-1$ and $+1$ respectively. This rescaling is only possible if the distribution of ideology scores is multi-modal, which is the case for each of the country-specific retweet networks in the current paper, but is not necesarily true for all social media interaction networks. 

\subsection*{Toxicity analysis}
Toxicity analysis is a common method in digital media research for identifying content which is ``rude, disrespectful, or unreasonable ... [and is] likely to make someone leave a discussion'' \cite{perspective}. 

There are a range of tools available for the automated detection of toxic content on social media. Here, we primarily use the Perspective API \cite{perspective}, developed by the Jigsaw team at Google, which provides toxicity scores between 0 and 1 corresponding to the probability that the classified content would be labeled as toxic by a human labeler. The Perspective API provides toxicity scores for English, German, French, Italian, Spanish, and Polish. The Perspective API cannot classify comments in Turkish. For Turkish language posts (and for English, French, Italian and Spanish language posts as a robustness check) we compute toxicity scores using Detoxify \cite{Detoxify}, developed by Unitary. In both cases, posts classified using these models do not require pre-cleaning. 

For the analysis of affective polarization, we compute the toxicity of all original tweets authored by a Twitter user with an assigned ideological score which mentions a single other user who also has an assigned ideological score for the same country. We only classify posts labeled as being in one of the seven languages covered by our toxicity analysis models. This does not include posts which only include URLs, or are authored in another language. We exclude tweets where multiple users are mentioned. Tweets analyzed include original posts, replies, and quote tweets. Importantly, we do not analyze the toxicity of retweets since the text of the retweet is attributable to the original author and not the retweeter. 

\subsection*{Statistical analysis}
Statistical analysis is carried out using the non-parametric Mann-Whitney U test. The test returns a $p$-value corresponding to the probability that the two distributions are drawn from the same parent distribution. Under the null hypothesis that the two samples are drawn from the same distribution, we use the standard convention that the null hypothesis can be rejected at three different significance levels: $p<0.05$: *, $p< 0.01$: **, $p<0.001$: ***. 

Point estimates for the median (mean) of the observables shown in Fig.~\ref{fig:fig3}, Fig.~\ref{fig:fig4}, and Fig.~\ref{fig:fig5} are computed using a bootstrapping procedure where the point estimate is sampled $1000$ times using a 50\% sample of the full distribution with replacement. The distribution of these point estimates is then shown as a boxplot for each country.

\subsection*{News media classification}
Media reliability data was provided by NewsGuard \cite{newsguard}. The data is proprietary and requires a license for use. The dataset by Newsguard includes a range of news media outlets in the US, UK, Canada, Germany, France and Italy from across the political spectrum and classifies the reliability of each outlet according to a set of journalistic criteria. Each outlet receives a score from 0 to 100 for each of the criteria, assigned by a team of independent journalists. Outlets with a score of 100 are considered the most reliable, whereas outlets with a score of 0 are considered the least reliable; lower scores reflect lower reliability. For the news media outlets classified, Newsguard list their online domains. From the tweets analyzed, we extract URLs and search for domains which correspond to a classified news domain. For each post which includes such a domain we assign the corresponding reliability score as provided by Newsguard. Analysis using media reliability scores provided by Newsguard have been shown to be similar to the results obtained using other media reliability datasets \cite{lin2023high}. 

\section*{Supplementary information}
Supplementary text and figures are provided online at \href{https://osf.io/wum3q/}{https://osf.io/wum3q/}

\section*{Acknowledgments}
M.F., F.Z., W.Q., and A.B. acknowledge the 100683EPID Project “Global Health Security Academic Research Coalition” SCH-00001-3391. M.F. thanks Maddalena Torricelli for help contextualizing the distribution of Italian politicians.

\bibliography{apssamp}

\end{document}